# Sequential dependencies between trials in free choice tasks

Carsten Allefeld[1,2*], Chun Siong Soon[1,2,4], Carsten Bogler[1,2], Jakob Heinzle[1,6], and John-Dylan Haynes[1,2,3*]

[1] Bernstein Center for Computational Neuroscience, Charité – Universitätsmedizin Berlin, Haus 6, Philippstrasse 13, 10115 Berlin, Germany; [2] Berlin Center for Advanced Neuroimaging, Charité – Universitätsmedizin Berlin, Charitéplatz 1, 10115 Berlin, Germany; [3] Berlin School of Mind and Brain, Humboldt-Universität zu Berlin, Luisenstraße 56, Haus 1, 10117 Berlin, Germany; [4] Neuroscience and Behavioral Disorders Program, Duke-NUS Graduate Medical School, 8 College Road, Singapore 169857, Singapore; [5] Department of Psychology, Technical University Dresden,Helmholtzstraße 10, 01069 Dresden, Germany; [6] Translational Neuromodeling Unit, Institute for Biomedical Engineering, University Zurich & ETH Zurich, Wilfriedstr. 6, 8032 Zurich, Switzerland

***Please direct correspondence to:**

Carsten Allefeld (carsten.allefeld@bccn-berlin.de) & John-Dylan Haynes (haynes@bccn-berlin.de)
Bernstein Center for Computational Neuroscience, Charité – Universitätsmedizin Berlin
Philippstrasse 13, Haus 6, 10117 Berlin / Germany

**Abstract**

In two previous experiments we investigated the neural precursors of subjects' "free" choices for one of two options (pressing one of two buttons, and choosing between adding and subtracting numbers). In these experiments the distribution of sequence lengths was taken as an approximate indicator of the randomness (or lack of sequential dependency) of the choice sequences. However, this method is limited in its ability to reveal sequential dependencies. Here we present a more detailed individual-subject analysis and conclude that despite of the presence of significant sequential dependencies the subjects' behavior still approximates randomness, as measured by an entropy rate (on pooled data) of 0.940 bit / trial and 0.965 bit / trial in the two experiments. We also provide the raw single-subject behavioral data.

## Introduction

Two previous experiments from our research group (Soon et al. 2008 and Soon et al. 2013 – in the following: "experiment 1" and "experiment 2") assessed the degree to which the outcome of subjectively free decisions can be predicted from preceding brain signals. Using a combination of functional magnetic resonance imaging and multivariate pattern classification, these experiments found that the outcomes of choices can be predicted with up to around 60 % accuracy from patterns of brain activity in medial prefrontal and parietal cortex as early as 7 seconds prior to the time when the participants believed to be making their decisions.

An interesting question in such a prediction of free choices is the degree to which a prediction might reflect or even be influenced by the choice on the previous trial (Lages and Jaworska 2012, Heinzle et al. 2009, Soon et al. 2008, 2013), especially since it is known that humans are particularly poor at generating random sequences (for a review see Nickerson 2002).

## Sequence length distributions

In the original papers (Soon et al. 2008, 2013) the assessment of randomness relied predominantly on the distribution of sequence lengths. This distribution is estimated by calculating the relative frequency with which sequences of exactly N identical choices occur in the time series, for each sequence length N. If a subject's choice sequence between two options A and B were say ABBAAABB this would mean one sequence with length N = 1, two sequences with length N = 2 and one sequence with length N = 3. Note that for the analysis of experimental data the first and last sequence from each run have to be discarded because it is impossible to tell whether they might have been cut off. In case the subject's choices are purely random, i.e. choices are sequentially independent and the two possibilities are chosen with the same probability, this is equivalent to the probability of sequences in tossing an unbiased coin. The resulting sequence length distribution is

$$p_N = 0.5^N.$$

Our original two papers reported that sequence length distributions showed a behavior very similar to this theoretical expectation. However, in these studies the data were pooled across subjects. In this report, we extend this information by the exact choice sequences for individual subjects and experimental runs (see Appendices 1 and 2). While the assessment of randomness based on sequence length distributions is limited, we include it here for comparison to the original papers before proceeding to more sensitive analyses.

Figure 1 shows the estimated probability of sequences with length N as a function of N for each individual subject S1–S12 (black bars) in experiment 1 (left or right hand button presses). For comparison, it also shows superimposed the expected distribution of sequence lengths for a process where each option is selected randomly and independently from the previous trial with a probability of 0.5 (red line). Please note that this parameter-free theoretical distribution neglects the effects of limited sampling, i.e. the influence of the limited number of trials per run. It fits the distributions for most subjects quite well, except for an overabundance of short trials. The same effect was apparent in the analysis of the data pooled across subjects (Supplementary Figure 2, Soon et al. 2008).

Figure 2 shows the same analysis for each individual subject S1–S17 in experiment 2 (adding or subtracting numbers). As in the previous experiment (Figure 1), the parameter-free theoretical function fits the distributions for most subjects quite well, again with the exception of a disproportional number of short trials, which was already apparent in the original analysis of pooled data (Supplementary Figure 1, from Soon et al. 2013).

## Probabilities of choices

One possible deviation from producing an ideal random sequence would be given if the subjects had a bias towards one of the two choices. The estimated probabilities (relative frequencies) of choosing "left" or "right" in experiment 1 are given in the columns $p_L$ and $p_R$ in Table 1, and for choosing "addition" or "subtraction" in experiment 2 in the columns $p_A$ and $p_S$ in Table 2, separately for each subject.

A statistical assessment of equiprobability is given by the p-value for a binomial test versus $p_L = p_R = 0.5$ and $p_A = p_S = 0.5$, respectively, in the column "binomial" of Tables 1 and 2. This application of the binomial test is not exact, because the test assumes independent samples, i.e. here sequential independence, which is not the case in this dataset (see below). To account for this fact, we assessed the p-values at a more lenient significance level of 0.1 to be more sensitive to possible deviations. The result is indicated by an asterisk in the tables: Equiprobability is formally rejected in 1 of 12 subjects in experiment 1, and 4 of 17 subjects in experiment 2. Please note that not rejecting the null hypothesis cannot be considered proof of its truth. Furthermore – for reasons of experimental design – in the original experiments participants had been pre-selected based on independent pilot experiments where the balance between both options was one criterion for selection (Soon et al. 2008). Thus the observed balance between left and right choices cannot be generalized to the full population of potential subjects.

The largest bias of 0.713 vs 0.287 occurs in S14 of experiment 2. On the whole, however, the choices were quite balanced in most subjects. Moreover, there is no systematic difference between choice probabilities across subjects; sign permutation tests applied to the probability differences ($p_L - p_R$ and $p_A - p_S$) result in p-values of 0.6 and 0.871 for experiments 1 and 2, respectively.

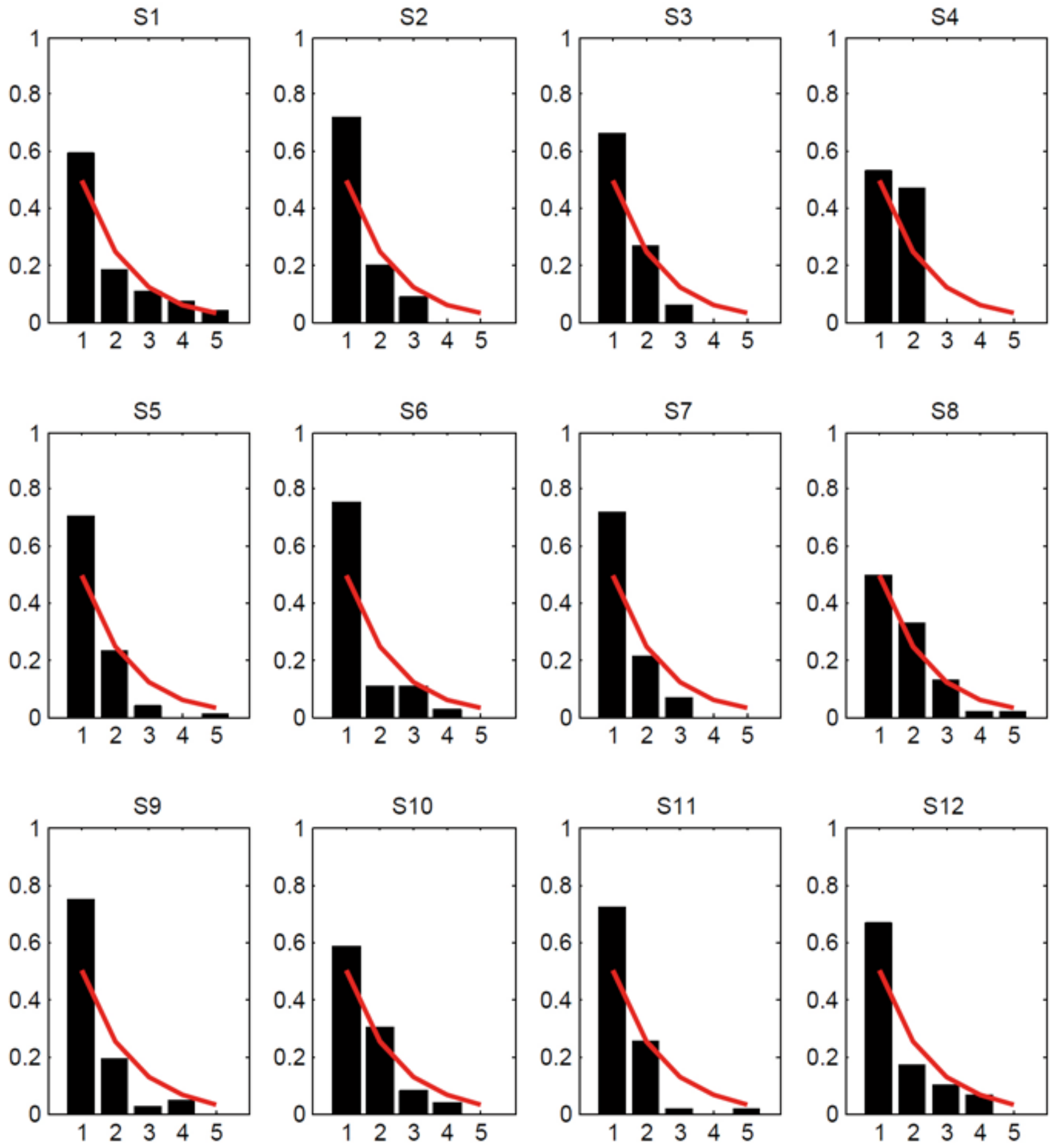

**Figure 1:** Distribution of sequence lengths from experiment 1, separately for the 12 subjects. The black bars show the probabilities estimated from the data and the red line the theoretical expectation for comparison.

## Prediction analysis

The previous two analyses show that the imbalance between choices is weak in most subjects and that the sequence lengths roughly follow the distribution that would be expected if choices were made sequentially independent and with equal probability. However, as expected from the fact that humans cannot produce perfectly random sequences (Nickerson 2002) small deviations from the expected distribution $p_N$ of sequence lengths were observed. We therefore assessed to which degree it would be possible to predict choice C(t + 1) from the previous choice C(t). In contrast to analyses of predictability of continuous data such as neurophysiological signals, each data point here can take on only two different discrete values. Therefore, instead of using a more complex classifier such as a support vector machine (see e.g. Müller et al. 2001 for a review), we used the following simple rule: We selected as the prediction of the next choice the one which followed most frequently the given previous choice.

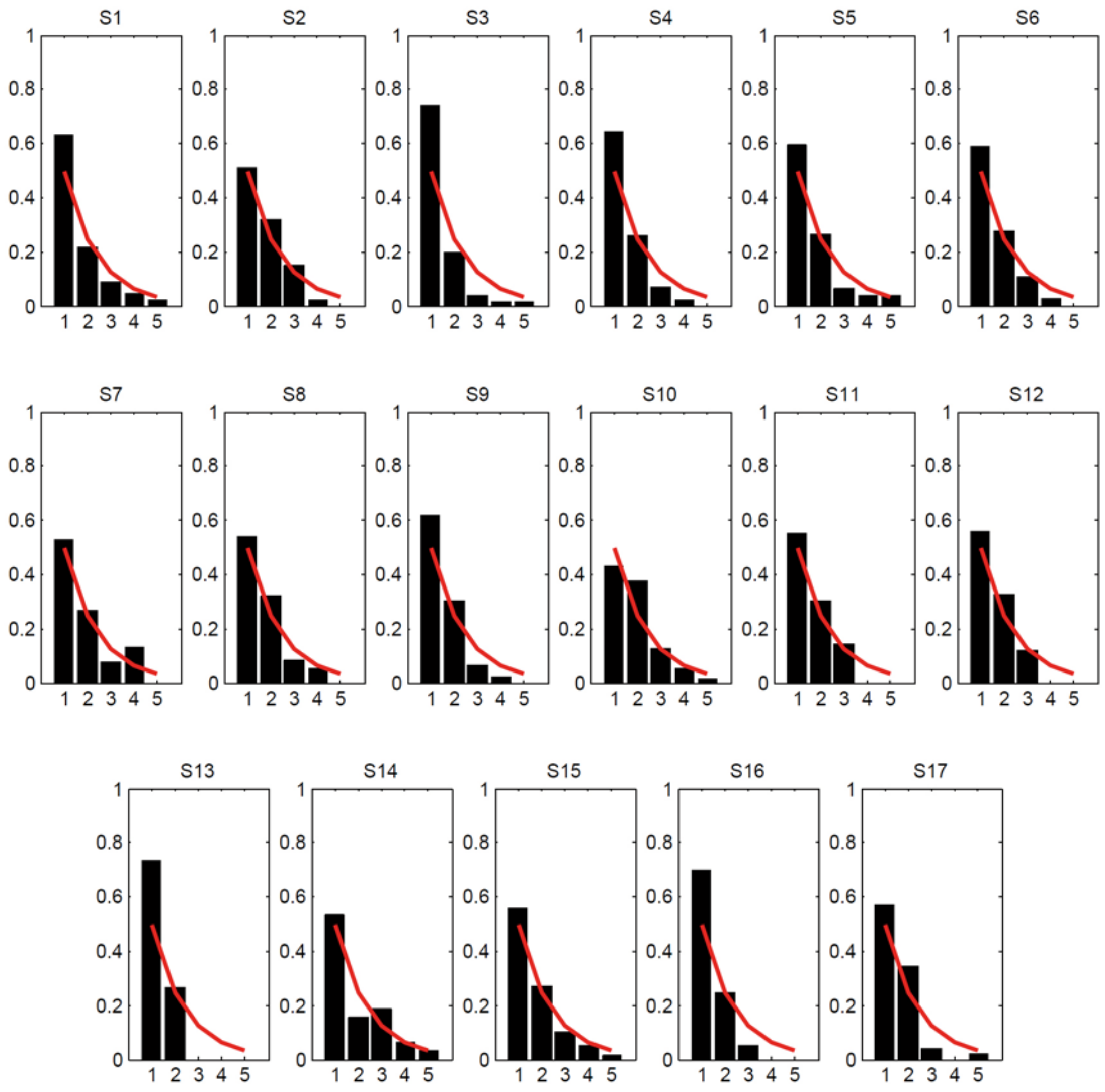

**Figure 2:** Distribution of sequence lengths from experiment 2, separately for the 17 subjects. The black bars show the probabilities estimated from the data and the red line the theoretical expectation for comparison.

That is, transition probabilities

$$T_{ij} = Pr(\ C(t + 1) = i \mid C(t) = j\ )$$

were estimated from the data based on how often the transitions occured in the data. For two choices, the four transition probabilities $T_{ij}$ are arranged in a $2 \times 2$ transition matrix T, whose entries add up to 1 in each column (see Figures 3 and 4 for examples). A transition matrix was estimated from a subset of the data ("training") and then used to predict choices in the remaining independent part of the data ("test"). Specifically, cross-validation was implemented across runs, meaning that the data from one run served as test data and the remaining runs served as training data, using each run in turn for testing. The resulting prediction accuracies were then averaged across cross-validation folds. Prediction of choice C(t+1) was implemented by selecting the column corresponding to the previous choice C(t) and then selecting the next choice depending on which of the two possibilities had the higher probability.

| | $p_L$ | $p_R$ | binomial | accuracy | Bernoulli | Markov | order | entropy | $p_{stay}$ | $p_{switch}$ |
|---|---|---|---|---|---|---|---|---|---|---|
| S1 | 0.528 | 0.472 | 0.6718 | 0.36 | 0.6541 | 0.6115 | 2 | 0.968 | 0.532 | 0.468 |
| S2 | 0.480 | 0.520 | 0.7644 | 0.62 | 0.0340 * | 0.8329 | 2 | 0.959 | 0.378 | 0.622 |
| S3 | 0.511 | 0.489 | 0.9179 | 0.69 | 0.0009 ** | 0.1906 | 3 | 0.870 | 0.310 | 0.690 |
| S4 | 0.449 | 0.551 | 0.4282 | 0.58 | 0.1492 | 0.1544 | 3 | 0.930 | 0.412 | 0.588 |
| S5 | 0.468 | 0.532 | 0.4976 | 0.71 | 0.0000 ** | 0.9453 | 1 | 0.873 | 0.287 | 0.713 |
| S6 | 0.600 | 0.400 | 0.0929 * | 0.66 | 0.0013 ** | 0.7741 | 2 | 0.838 | 0.329 | 0.671 |
| S7 | 0.543 | 0.457 | 0.4351 | 0.75 | 0.0000 ** | 0.3847 | 2 | 0.813 | 0.253 | 0.747 |
| S8 | 0.476 | 0.524 | 0.6536 | 0.57 | 0.1957 | 0.0175 * | 2 | 0.926 | 0.439 | 0.561 |
| S9 | 0.461 | 0.539 | 0.5250 | 0.74 | 0.0000 ** | 0.3158 | 3 | 0.841 | 0.266 | 0.734 |
| S10 | 0.522 | 0.478 | 0.7069 | 0.59 | 0.1132 | 0.5231 | 3 | 0.974 | 0.417 | 0.583 |
| S11 | 0.472 | 0.528 | 0.5887 | 0.74 | 0.0000 ** | 0.0265 * | 2 | 0.790 | 0.274 | 0.726 |
| S12 | 0.533 | 0.467 | 0.6445 | 0.62 | 0.0807 | 0.7939 | 2 | 0.952 | 0.385 | 0.615 |
| | 0.500 | 0.500 | | | | | | 0.940 | 0.353 | 0.647 |

**Table 1**: Analysis of behavioral data from experiment 1. For 12 subjects, the columns show: estimated probabilities for choices "L" and "R"; p-value and significance ($\alpha$ = 0.1) of a binomial test versus $p_L = p_R = 0.5$; cross-validated accuracy for predicting a choice from the previous one; p-value and significance ($\alpha$ = 0.05, 0.01) for a test of sequential independence (Bernoulli process); p-value and significance ($\alpha$ = 0.05, 0.01) for a test of sequential dependence on the last choice only (Markov property); estimated optimal order of a Markov model; entropy rate in bit / trial of the choice sequence according to a second-order Markov model; estimated probabilities for staying with the same choice or switching to the other choice. Some of the values are also given with respect to data pooled across subjects in the last row.

The cross-validated prediction accuracies for each subject are shown in the column "accuracy" in Tables 1 and 2. The average classification accuracy across subjects is 0.64 (±0.031) in experiment 1 and 0.62 (±0.015) in experiment 2.

| | $p_A$ | $p_S$ | binomial | accuracy | Bernoulli | Markov | order | Entropy | $p_{stay}$ | $p_{switch}$ |
|---|---|---|---|---|---|---|---|---|---|---|
| S1 | 0.542 | 0.458 | 0.4394 | 0.58 | 0.1548 | 0.9067 | 2 | 0.976 | 0.423 | 0.577 |
| S2 | 0.495 | 0.505 | 1.0000 | 0.62 | 0.0379 * | 0.0359 * | 3 | 0.912 | 0.387 | 0.613 |
| S3 | 0.532 | 0.468 | 0.4976 | 0.70 | 0.0000 ** | 0.9161 | 3 | 0.855 | 0.295 | 0.705 |
| S4 | 0.538 | 0.462 | 0.3588 | 0.66 | 0.0001 ** | 0.3659 | 3 | 0.926 | 0.342 | 0.658 |
| S5 | 0.482 | 0.518 | 0.7750 | 0.58 | 0.0750 | 0.4165 | 2 | 0.956 | 0.410 | 0.590 |
| S6 | 0.572 | 0.428 | 0.0964 * | 0.62 | 0.0009 ** | 0.1561 | 2 | 0.906 | 0.370 | 0.630 |
| S7 | 0.474 | 0.526 | 0.6398 | 0.48 | 0.5560 | 0.2925 | 2 | 0.980 | 0.538 | 0.462 |
| S8 | 0.539 | 0.461 | 0.4264 | 0.58 | 0.0671 | 0.3290 | 2 | 0.958 | 0.415 | 0.585 |
| S9 | 0.489 | 0.511 | 0.8250 | 0.67 | 0.0000 ** | 0.0368 * | 2 | 0.877 | 0.328 | 0.672 |
| S10 | 0.618 | 0.382 | 0.0044 * | 0.61 | 0.8604 | 0.0191 * | 2 | 0.919 | 0.535 | 0.465 |
| S11 | 0.500 | 0.500 | 1.0727 | 0.60 | 0.0560 | 0.0136 * | 2 | 0.895 | 0.400 | 0.600 |
| S12 | 0.500 | 0.500 | 1.0858 | 0.58 | 0.2506 | 0.0665 | 3 | 0.904 | 0.421 | 0.579 |
| S13 | 0.482 | 0.518 | 0.8939 | 0.63 | 0.1392 | 0.2991 | 2 | 0.870 | 0.370 | 0.630 |
| S14 | 0.713 | 0.287 | 0.0000 * | 0.73 | 0.3048 | 0.0301 * | 3 | 0.777 | 0.538 | 0.462 |
| S15 | 0.492 | 0.508 | 0.9307 | 0.57 | 0.1501 | 0.2925 | 2 | 0.970 | 0.434 | 0.566 |
| S16 | 0.578 | 0.422 | 0.1371 | 0.72 | 0.0000 ** | 0.0430 * | 2 | 0.767 | 0.283 | 0.717 |
| S17 | 0.336 | 0.664 | 0.0004 * | 0.66 | 0.0315 * | 0.2593 | 3 | 0.884 | 0.459 | 0.541 |
| | 0.524 | 0.476 | | | | | | 0.965 | 0.405 | 0.595 |

**Table 2:** Analysis of behavioral data from experiment 2. For 17 subjects, the columns show the estimated probabilities for choices “A” and “S”; for an explanation of the further columns, see Table 1.

## Stochastic process analysis

In agreement with previous findings (Nickerson 2002), the sequence length statistics and the prediction analyses indicate that there are weak deviations from purely random behavior in the choice sequences of subjects.

The existence of such dependencies over one step can be assessed exactly for each subject by testing the null hypothesis of a Bernoulli process (using Fisher's exact test applied to the transition counts treated as a contingency table; compare Fisher 1954). The resulting p-values are shown in Tables 1 and 2, column “Bernoulli”. The null hypothesis of no sequential dependency over one step can be rejected at the standard significance level of 0.05 for 7 of 12 subjects in experiment 1, and for 7 of 17 subjects in experiment 2; in most of these the p-value is even below 0.01.

There are two natural follow-up questions: The first is how strong the dependencies are. The second is whether sequential dependencies are limited to one step, i.e. whether the choice behavior of the subjects conforms to a first-order Markov process.

We first turn to the question of Markov order. The null hypothesis of a first-order Markov process can be tested using the “minimum discrimination information” approach of Kullback et al. (1962); the resulting p-values are shown in Tables 1 and 2, column “Markov”. According to this test, sequential dependencies exceeding one step are present in 2 of 12 subjects in experiment 1, and 5 of 17 subjects in experiment 2.

Hypothesis tests have a bias in favor of retaining the null hypothesis, and even more so if applied in sequence to several nested hypotheses. A better way to assess the extent of sequential dependencies is to determine which order a Markov model needs to have such that it optimally describes the data. Optimal model orders estimated according to the approach of Csiszár and Shields (2000) based on the Bayesian information criterion are shown in Tables 1 and 2, column “order”. Data from most subjects appear to be optimally modelled by a second-order Markov process.

Next we turn to the strength of the dependencies. The estimated Markov model order assesses the temporal extent of sequential dependencies, but does not quantify their strength, or the degree of randomness that the sequential choice behavior of subjects retains despite of the influence of previous choices on the next one. The standard way to measure randomness is information-theoretic entropy, which quantifies the amount of information (surprise) an observer receives when notified of the outcome of a random event (see Cover and Thomas 1991). Its adaptation to the case of stochastic processes is the entropy rate, the average amount of information received per time step from observing a sequence of events. The entropy rate of the choice behavior of subjects, estimated according to a second-order Markov model, is given in Tables 1 and 2 for each subject in experiment 1 and 2, respectively. Single-subject values vary from 0.767 to 0.980 bit / trial; estimated on data pooled across subjects, the entropy rate is 0.949 bit / trial in experiment 1 and 0.965 bit / trial in experiment 2. For comparison, in the purely random case of independent equal-probability choices between two alternatives, the entropy rate would be 1 bit / trial.

## The sequential choice behavior in detail

How did the subjects’ choice behavior deviate from randomness? The left panels of Figures 3 and 4 show one-step transition matrices estimated from the pooled data of all subjects in experiment 1 and 2, respectively. They suggest that transitions occur more often in the direction of switching to the other choice (“L” → ”R”, “R” → ”L” and “A” → ”S”, “S” → ”A”) than staying with the same choice, in a proportion of about 0.6 vs 0.4 on average. Stay and switch probabilities estimated for each subject separately, shown in Tables 1 and 2 in columns $p_{stay}$ and $p_{switch}$, confirm this effect to systematically occur across subjects. Estimated switch probabilities are higher than stay probabilities in 11 out of 12 subjects in experiment 1, and 14 out of 17 subjects in experiment 2. This tendency to switch also accounts for the higher frequency of sequences of length 1 (see above) than what would be expected from a purely random process.

The right panels of Figure 3 and 4 show the probabilities for the next choice depending on the *last two* previous choices, representing Markov models of order 2. They suggest that the preference to switch is more generally a tendency to avoid longer sequences of the same choice. Especially in experiment 2 (Figure 4), the probability to switch to the other choice is even higher if the two previous choices had been identical, and is lower if there was already a switch immediately before.

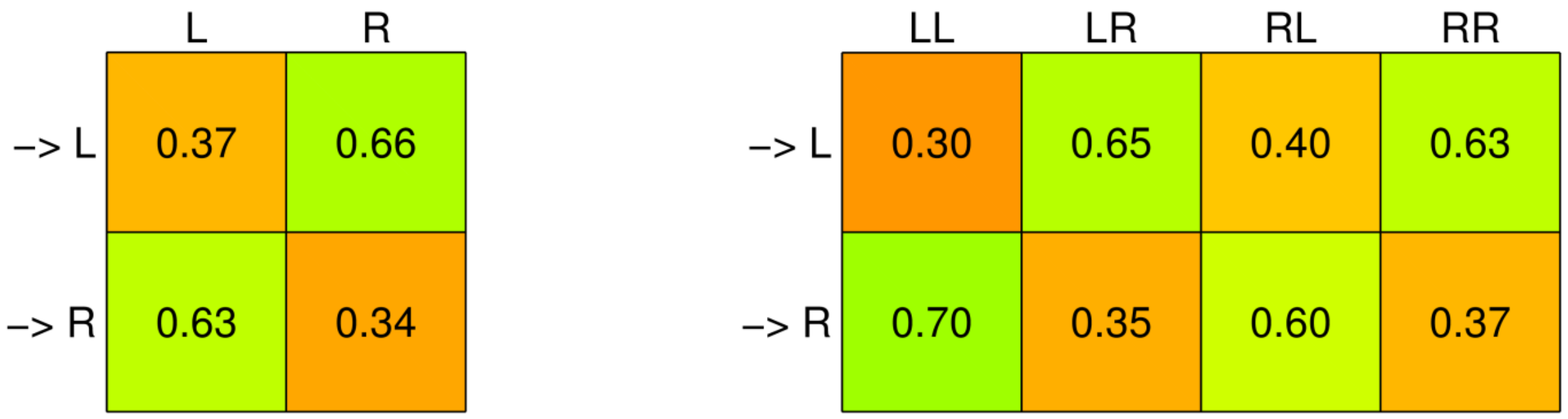

**Figure 3:** Transition probabilities estimated from the data of experiment 1. Left panel: First-order Markov model. The two columns correspond to the two possible previous choices, the rows to the next choice. Right panel: Second-order Markov model. The four columns correspond to the four possible pairs of the two previous choices. Probabilities add up to 1 for each column separately.

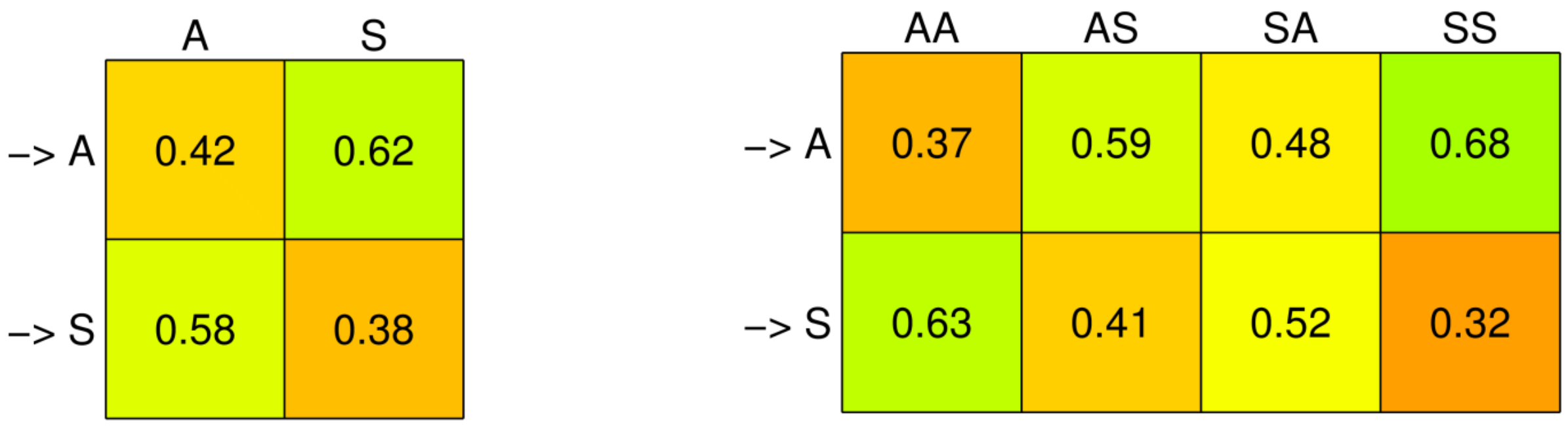

**Figure 4:** Transition probabilities estimated from the data of experiment 2.

## Discussion

In this brief report, we analyzed the statistical properties of the sequential choice behavior of subjects in two previous experiments (Soon et al. 2008, 2013) in more detail and at the individual subject level.

In most subjects, both possibilities were chosen approximately equally often. This is not surprising given that subjects were selected for participation in the experiments based on whether they exhibited an approximate balance between choices in a pilot experiment before scanning (see Supplementary Material of Soon et al. 2008). The sequence length distributions in many subjects follow roughly a distribution that would be expected if choices are equiprobable and sequentially independent, but with a higher frequency of sequences of length 1. The underlying sequential dependencies can be utilized to predict the next choice with an accuracy of about 60 %, and can be shown to be present in most subjects. Analysis of the data using a Markov model approach indicates that dependencies extend across about two previous choices, but lead only to a small reduction in the amount of randomness as quantified by the entropy rate (ca. 0.95 bit / trial). An examination of the transition probabilities shows that subjects actively avoid longer sequences of identical choices, which accounts for the findings of the sequence length statistics.

The weak behavioral predictability observed here has been previously reported (Nickerson 2002, Lages & Jaworska 2012) and has also been observed previously in our own work when using more sensitive dependency measures other than sequence length distributions (Heinzle, Usnich & Haynes 2009). One interesting question is how this dependency is related to brain-based prediction of choices (Lages & Jaworska 2012). Please note that the behavioral prediction accuracy cannot be directly compared to the published brain-based prediction accuracy of choices because the latter is based on *aggregate* brain measures (runwise parameter estimates rather than single trials, for details see Soon et al. 2008, 2013). Importantly, there are several reasons that speak *for* and *against* sequential dependencies as a cause of early predictive information (see Haynes 2011 and Lages & Jaworska 2012 for discussions). Furthermore, besides sequential dependencies there are several possibilities why early brain activity might be predictive of upcoming choices (Haynes 2011). One way to assess the influence of previous trials on brain-based decoding accuracy is to perform decoding using labels shifted by one trial. A subsequent paper will present a reanalysis of neuroimaging data to address this question.

# Appendix

## Appendix 1: Behavioral choices of Experiment 1

Sequences of choices between left and right button presses in 10 runs for each of the 12 subjects whose fMRI entered the final analysis.

**Subject 1**
RLRLRLLLRL
LRRRRR
RRRRLLLL
LLLRRRRLR
RRLLLLLLL
RRLRRRRLLR
RLLLLLRLRL
RRLLLRLRL
LRLLLRRLRL
LRRLLRRL

**Subject 2**
RLRRRLLRLRR
RLRLRLLLL
LRLRLRLRRL
RRLRLLRLRR
RLRRRLLRLR
RRRRRLLLRL
RRRLRLLLRR
RLRLRRLRLRL
RLLRLLRLLL
LRRLRRLLL

**Subject 3**
RLRLLRLLR
RRLLRLRRRL
RLRLRLLRR
LRLRRLRRRL
LRRLLRLRLR
RLLRRLLRLL
LRLRLRLLRL
RRLLRLLLL
LRRRLRLLR
RLRLRLRL

**Subject 4**
RLRLRRRR
RRRRLLR
RLLRRLRR
RLLRRLRR
RRRLRRL
RLRRLLRL
RLRLLRRL
LRRLLLL
RLLRRLLRL
LLRLRLRL

**Subject 5**
RLRRLLRLLLLLR
LLRLRLRLLRRLR
LRRRLRLRRLLRRR
RLLRLRLRLRRLLR
RLRLRLLRLLLRL
LRRLRRLRLRLRLR
LRLLRRLRLRLRRLR
RLRRLRLLRLRRRLRL
LRLRLRLRRLRLR
RLRRLRRRRRRLRL

**Subject 6**
RLLLRRLL
LLRLRLLR
LRLRLLLR
LRLRLRLRLL
RLRLRRRLRL
RRLLLLR
LRLLRRLL
LRLRLRLL
RLLLRLL
LLRLRL

**Subject 7**
RLLRLRLLR
RLRLLRLRL
RLRLLRLRL
LRRLLRRRLLR
LRLRLRLLLRR
RLLLRLLRLRLL
RLRLRLLLRRL
LRLRLRRLRRL
LRLRLLRLRLR
LLRLRRLRLRL

**Subject 8**
RLLRLRRRRRLR
RLRRLRRLRRRL
RRLLRLLRRRLLR
RLLRRLRLLLRR
LLRRLLLRLRRLL
RRRLLRLRRLLLR
RLLLRLRLLRRL
LRLRLLRLRRR
LLRLRLLRLLRR
LLRRRRLRLLRRRL

**Subject 9**
RLRLRLR
LRRRRLLR
RLRRRRLRR
LRRLRRRLLRLR
RLRLRLRLRL
RLRLRRLLRL
RLLRLRLLR
RLRLRLLR
RLRLRLLL
RRLLRLRL

**Subject 10**
RLRLLRLLLLR
LRLRLRLLRLL
RRRRLLLLRRRLL
RLLRLLRRRLRR
LLLLLRLRLLRLRL
RLRLLRLRRLL
LLLRRRLLRRL
RLRLLRRLRR
RLRRLRRLLR
RLRRRLRLLR

**Subject 11**
RLRLRLRRLRLLRLL
RLLRRLRLRLRLRLRR
RLRRRRRLLRLRLRR
RLLRRLRRRLLRL
RRLLRLLRLRLLRL
LLRRLRLRLRRLR
RRLRRLLRRL
RLRLRLRLL
RLRLRLRLR
LRLLRRLRL

**Subject 12**
LLRRLRLL
LRLLRLRR
LLRRRLR
LRLLLRL
RLLRLLLLRL
RLRRRRLR
RLRRRLL
LLLRRLR
LLRLRLR
RLLRLR

## Appendix 2: Behavioral choices of Experiment 2

Sequences of choices between “adding” and “subtracting” in 10 runs for each of the 17 subjects whose fMRI entered the final analysis.

**Subject 1**
AASASASSSAS
ASAASSSSSAA
ASASSASSA
ASAAAASS
AAASAASASAS
AAASAAASSASSA
SSAASASAA
ASASASAAAAS
ASAAASSASSS
AASSSASAASSAA

**Subject 2**
SAASAAASSA
SSAAASSASAA
ASSSAASSA
ASASASSSA
AASSASASA
ASSSSASAAAS
ASSASAAASSA
ASSASSAASAS
SSASAAASSAS
ASSAASSASAS

**Subject 3**
AASSSASSAAAAAS
ASSASASAAASAS
SSASASAASSASAS
SSAASSASAASSSA
ASAASSSSASA
SASASASASAAAA
ASSASASASASASS
ASAASASASASAASA
AASSASAASASAASA
ASASAASAASASASAA

**Subject 4**
ASAASASASAAAAAA
ASAASASSAASASAAS
SAAASASASSSA
ASAAASASASSAASAS
ASASAASAASAASSSAA
ASASSAASASSAASAAS
ASAAASAASASSSSAASS
AASAAAASASASSAASASSS
ASSASAASSASASSSASASA
ASASASSAASASSSASASSA

**Subject 5**
AASASSA
ASSSSSAASA
SSASSSSAAA
ASAAASSAS
ASSSASSA
AASASSAASS
ASASASSASAAS
SSAAAAASASA
SASASASSASAAAASSAS
SSASAASSAASSSAS

**Subject 6**
ASSAASAAASSSA
SAAAASSAAAAAA
ASASAASASASASAS
AASAASSSAASASA
SAASASAASAASSSA
SAASASSAASAASSA
SSAASAASAAASAAS
ASASAASASASSAS
ASAAASASAAASSAS
ASAASASSSAAAASAS

**Subject 7**
AASAAASASSS
SSSASAASSS
SSASAAASAS
SSSSSAASAAS
SAASSSSAS
SSSSAASASSAA
ASAASSSSASAA
AASSSAASSAAAA
SSAAAASSASAAA
ASAAAASSSSASS

**Subject 8**
AASSAASAS
ASSAASSAAA
ASSAASAASASSAS
AAASSSAAAASA
ASSSAAASASAAS
SAASSASSAASSA
SSASASASAAAA
ASASSSAAAASSSSASA
ASAAASASAASASA
SSSAASAASASSAS

**Subject 9**
AASASSASAASASS
SAAASSAASSAASSASSASA
AASASSASSAASSAASA
SASSASASAAAASSASAS
AASSSASAASAASASAS
SSAASASSAASASASASA
ASSAAASSSSAAASSSAAS
SASSAASSASASASSSASAS
SAASAASASSSAASASSASAS
SASASASASASAASASSAAS

**Subject 10**
ASSAASASSSASAAA
SSAASSAAAAAAAAAA
AAASAASAASAAAAS
AAAASAAAAASAA
AASAAASAAASSAAA
SAAAASSASSAASSSAS
ASSASAAASSAAAS
ASSAASAASSASAAA
AAAASASSSSASSAA
ASAAASSASSAASSASS

**Subject 11**
SSASSAASAA
SSAAASSSASA
SSAAASAASSAA
SAASSAASSSAAS
SASASSAAASSSAA
AASASSASSSAA
SSASASASASSAASA
ASASSASASASS
AASAASAAS
ASASSAAASSAS

**Subject 12**
AASAASSSA
SSAASSASA
AAASSSAS
AASAASSS
AAASASA
ASAASASSS
SSASSAAS
SASASSAA
ASSSAASASA
ASSSAASAAS

**Subject 13**
AASASAS
SSAASS
SSASAS
SSASAA
ASSAASS
AAASA
ASAA
SAASA
SASSS
ASSAS

**Subject 14**
SAAAAAASASS
SASAAASSAA
AASAASAAASA
AAAASAAAASS
ASAAAAASA
SAAASAASSA
AASAAASAAA
ASSAAAASAA
ASAAASAAA
SAAASAAAAA

**Subject 15**
SAAASSSSSAA
AASSASASSASSA
ASSSSAAAASASSA
ASAASASSSAAAA
AAAASSSASASSSSA
AASSASAAASASA
ASASAASASSAA
ASASASSASSAA
SSASSSASSASSAAS
SAASASSSAASSAS

**Subject 16**
AASAASAS
ASAASASSA
ASAAASAASA
AASAASASASS
SASAAASASA
ASAASAASAS
SAAASSAASAAS
ASASAASASAAA
SASAASASAS
SASAASSASS

**Subject 17**
ASSASSSSSASSA
SAASSAASSASSA
SSSSAASASA
ASAASSSASASS
SSAASASSASSAS
SSSSSSASAAAS
SSASSASSASSSS
SASSSSSSA
ASSASASSASSS
SSSSSASASSAS